\documentclass[twocolumn,prl,showpacs,amsmath,amssymb,superscriptaddress,floatfix]{revtex4}

\usepackage[dvipdfmx]{graphicx}
\usepackage{dcolumn}
\usepackage{bm}
\usepackage{color}

\begin{document}

\title{
Nature of isomerism in exotic sulfur isotopes 
}

\author{Yutaka Utsuno}
\email{utsuno.yutaka@jaea.go.jp}
\affiliation{Advanced Science Research Center, Japan Atomic Energy
Agency, Tokai, Ibaraki 319-1195, Japan}
\affiliation{Center for Nuclear Study, University of Tokyo, Hongo,
Bunkyo-ku, Tokyo 113-0033, Japan}

\author{Noritaka Shimizu}
\affiliation{Center for Nuclear Study, University of Tokyo, Hongo,
Bunkyo-ku, Tokyo 113-0033, Japan}

\author{Takaharu Otsuka}
\affiliation{Department of Physics, University of Tokyo, Hongo,
Bunkyo-ku, Tokyo 113-0033, Japan}
\affiliation{Center for Nuclear Study, University of Tokyo, Hongo,
Bunkyo-ku, Tokyo 113-0033, Japan}
\affiliation{National Superconducting Cyclotron Laboratory, 
Michigan State University, East Lansing MI 48824, USA}

\author{Tooru Yoshida}
\affiliation{Center for Nuclear Study, University of Tokyo, Hongo,
Bunkyo-ku, Tokyo 113-0033, Japan}

\author{Yusuke Tsunoda}
\affiliation{Department of Physics, University of Tokyo, Hongo,
Bunkyo-ku, Tokyo 113-0033, Japan}
\date{\today}

\begin{abstract}

We clarify the origin of the anomalously hindered $E2$ decay 
from the $4^+_1$ level in $^{44}$S by   
performing a novel many-body analysis in the shell model. 
Within a unified picture about the occurrence of isomerism 
in neutron-rich sulfur isotopes, 
the $4^+_1$ state is demonstrated to be a $K=4$ isomer 
dominated by the two-quasiparticle configuration 
$\nu\Omega^{\pi}=1/2^-\otimes\nu\Omega^{\pi}=7/2^-$. 
The $4^+_1$ state in $^{44}$S is a new type of high-$K$ 
isomer which has significant triaxiality. 
\end{abstract}


\maketitle

Among the most fundamental properties of nuclei is 
quadrupole collectivity, based on which the yrast $0^+$, $2^+$, 
$4^+$, $\ldots$ states in even-even nuclei are in general connected 
with strong $E2$ matrix elements. 
For open-shell nuclei in particular, $B(E2)$ values 
between neighboring yrast states are tens to hundreds of times larger 
than the Weisskopf estimate \cite{bohr}.
Contrary to this common sense, a recent experiment has reported 
\cite{s44isomer} that 
the $4^+$ state of $^{44}$S located at 2.4 MeV, 
most likely the yrast state due to its small
$E_x(4^+)$ to $E_x(2^+_1)$ ratio 1.9, 
has a strongly hindered $B(E2)$ 
value ($\lesssim 1$~W.u.) for the transition to $2^+_1$. 
While this quite unusual 
$E2$ property of the $4^+_1$ state, a kind of isomer,   
has been described with shell-model calculations \cite{s44isomer}, 
its underlying nuclear structure and lowering mechanism 
are still unclear. 
While a $K=4$ high-$K$ isomeric state is suggested 
in analogy to heavy-mass nuclei \cite{s44isomer}, 
this hypothesis is not supported by later microscopic 
calculations \cite{s44gogny, s44qinv}.

Besides the $4^+_1$ level in $^{44}$S, 
plenty of exotic nuclear properties have been reported for neutron-rich 
sulfur isotopes. 
A modest $B(E2;0^+_1 \to 2^+_1)$ value in $^{44}$S \cite{s44be2} 
indicates the development of quadrupole collectivity 
despite the neutron magic number 28. 
An extraordinary low-lying isomeric $0^+_2$ state in $^{44}$S 
\cite{s44zero1, s44zero2} 
might suggest a spherical-deformed shape coexistence. 
Similarly, an isomeric $7/2^-_1$ state in $^{43}$S 
is also possibly an indication of shape coexistence 
\cite{s43isomer, s43g}. 
Those observations have triggered 
state-of-the-art theoretical investigations 
based on the large-scale shell-model calculations 
\cite{sdpf-u, n28sm, epqqm, sdpf-mu, s44qinv}, 
the beyond-mean-field approaches \cite{peru, n28gogny, s44gogny, ddpc1}, 
and the antisymmetrized molecular dynamics (AMD) \cite{amd_s}. 

In this Letter, we demonstrate that the isomeric $4^+_1$ state 
in $^{44}$S occurs due to the dominance of a $K=4$ intrinsic state 
by means of beyond-mean-field approximations to the shell model. 
We also present a unified understanding of the occurrence of the exotic isomers 
in neutron-rich sulfur isotopes $^{43,44}$S, 
thus confirming the robustness of the present approaches and results.
$^{44}$S is the lightest-mass case among the high-$K$ isomers ever
identified 
in the $A\sim 100$, $A\sim 130$, $A\sim 180$, 
and $A\sim 250$ regions \cite{isomer_chart}.
More surprisingly, this nucleus is triaxially deformed in contrast 
to the known cases having well-developed axially
symmetric shapes \cite{isomer_review}. 

We start with the conventional 
shell-model calculations for neutron-rich nuclei around $N=28$ 
in the $\pi (sd)^{Z-8} \nu (pf)^{N-20}$ valence space 
with the SDPF-MU interaction \cite{sdpf-mu}. 
As we will show in more detail later, 
the nuclear structure of the sulfur isotopes 
of the present interest is very well reproduced with 
the SDPF-MU interaction,  
as well as with the SDPF-U interaction \cite{sdpf-u}. 

While the shell-model calculation is capable of 
describing observables of nuclei quite quantitatively, 
it is not necessarily easy to draw a comprehensive picture of 
nuclear structure, 
in particular from the intrinsic-frame point of view.
Introducing an appropriate mean-field based method into the shell model 
is a key to solving this problem. 
The method to be taken should 
represent spin-dependent intrinsic
structure to describe the abrupt change between $2^+_1$ and $4^+_1$ in
$^{44}$S and should
provide high-quality many-body wave functions  
comparable to the full shell-model calculation.
Clearly, simple mean-field calculations such as the Hartree-Fock method 
cannot satisfy those demands.
Here we take the variation after angular-momentum projection (AM-VAP) 
as a beyond-mean-field method to efficiently 
describe spin dependence
within a framework that can 
well define an intrinsic state.

In the AM-VAP, wave functions are 
determined to minimize the energy 
$E(I\sigma) = {\langle IM\sigma | H | IM\sigma \rangle_{\rm AM-VAP}}/
{\langle IM\sigma  | IM\sigma \rangle_{\rm AM-VAP}}$ 
 in the form of wave function 
\begin{equation}
\label{eq:amvap}
| IM\sigma \rangle_{\rm AM-VAP} = \sum_K g^{IM\sigma}_K \hat{P}^I_{MK} |  
\Phi(IM\sigma)  \rangle,   
\end{equation}
where $| \Phi(IM\sigma)\rangle = 
\prod_k\left( \sum_l D_{lk}^{IM\sigma}c^{\dag}_l\right)|-\rangle
$ is a general single Slater determinant
parametrized by $D_{lk}^{IM\sigma}$ 
and can be regarded as the intrinsic state of 
$| IM\sigma \rangle_{\rm AM-VAP}$.
$\hat{P}^I_{MK}$ 
is the usual angular-momentum projection operator \cite{ring}
with $I$, $M$ and $K$ denoting the total angular momentum and 
its $z$ components along the laboratory and intrinsic frames, 
respectively, and each state with a given $(I, M)$ is labeled with
$\sigma$.  
The mixing of $K$ in $| IM\sigma \rangle_{\rm AM-VAP}$ is 
represented by $g^{IM\sigma}_K$. 
$D_{lk}^{IM\sigma}$ and $g^{IM\sigma}_K$ are the variational parameters 
which are optimized using the conjugate gradient method \cite{numerical}
(see \cite{rmcsm} for the expression of 
the gradient of $E(I\sigma)$).
Furthermore, we calculate the overlap probabilities between the solutions 
found in this procedure and the shell-model wave functions of interest 
in order not to miss important solutions. Those overlap probabilities 
also work to test the quality of the AM-VAP wave functions. 
Multiple AM-VAP solutions, if found, are treated as separate levels 
if they are nearly orthogonal.

\begin{figure}[bt]
 \begin{center}
 \includegraphics[width=8.0cm,clip]{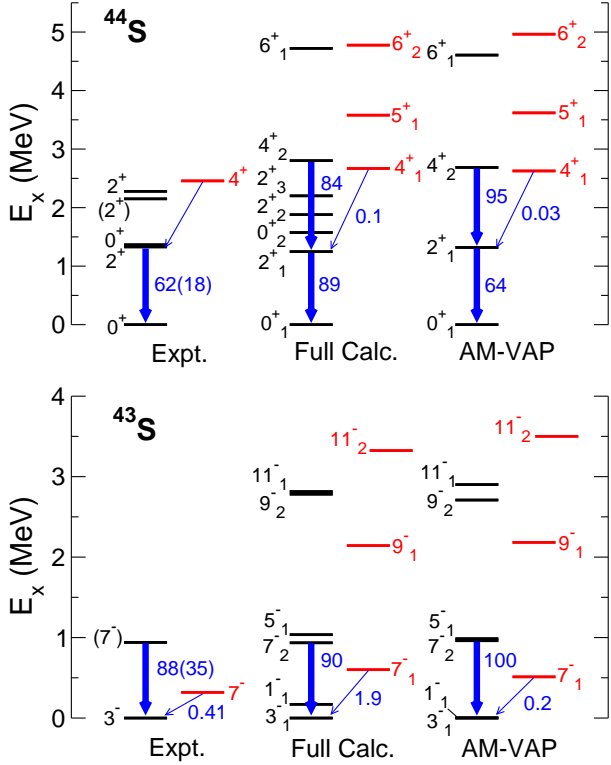}
 \caption{
(Color online) Energy levels 
and $B(E2)$ values (in $e^2$fm$^4$) in $^{44,43}$S compared among 
experiments (Expt.), the full shell-model calculation (Full Calc.), and 
the AM-VAP approximation to the shell model (AM-VAP). 
The spin-parity is denoted as $J^{\pi}$ for $^{44}$S and $2J^{\pi}$ for
  $^{43}$S.  Experimental data are taken from 
\cite{s44isomer, s44be2, s44zero1, s43g, s44gamma, s43e2}.
}
 \label{fig:s-level}
 \end{center}
\end{figure}

Figure~\ref{fig:s-level} shows energy levels 
and $B(E2)$ values in $^{43,44}$S   
compared among experiments, the full shell-model calculation, 
and the AM-VAP calculation. 
The $E2$ effective charges $(e_p, e_n)=(1.35e, 0.35e)$ are taken. 
As presented in Fig.~\ref{fig:s-level}, 
the results of the AM-VAP calculations 
are very close to 
those of the full shell-model calculations, including a strongly hindered
$B(E2; 4^+_1 \to 2^+_1)$ value in $^{44}$S. 
The overlap probabilities calculated between the full shell-model states 
and the AM-VAP states are rather close to unity: 
0.92, 0.81, 0.86, 0.88, and 0.90 
for the $0^+_1$, $2^+_1$, $4^+_1$, $4^+_2$, and $5^+_1$ states 
in $^{44}$S, respectively, and 
0.96, 0.91, 0.85, 0.93, 0.93, 0.92, 0.93, 0.92, and 0.84 for 
the $1/2^-_1$, $3/2^-_1$, $5/2^-_1$, $7/2^-_1$, $7/2^-_2$, 
$9/2^-_1$, $9/2^-_2$, $11/2^-_1$, and $11/2^-_2$ states in $^{43}$S, 
respectively. 
The $6^+$ states of the AM-VAP split into the two 
shell-model states because of the accidental degeneracy of the $6^+$ states 
in the shell-model calculation. 

\begin{table}[t]
\caption{Distribution of $|K|$ and deformation parameters 
$(\beta,\gamma)$ ($\gamma$: in deg.) in the AM-VAP states for $^{44}$S. }
\label{tab:s44}
\begin{ruledtabular}
\begin{tabular}{rrrrrrrrrr}
$I^{\pi}_{\sigma}$ & \multicolumn{7}{c}{$|K|$} & $\beta$ & $\gamma$ \\ \cline{2-8}
& 0 & 1 & 2 & 3 & 4 & 5 & 6 &  &  \\
\hline
$0^+_1$ & 1.00 & & & & & & & 0.24 & 33\\
$2^+_1$ & 0.98 & 0.00 & 0.01 & & & & & 0.26 & 23 \\
$4^+_2$ & 0.92 & 0.08 & 0.00 & 0.00 & 0.00 & & & 0.28 & 14 \\
$6^+_1$ & 0.76 & 0.23 & 0.01 & 0.00 & 0.00 & 0.00 & 0.00 & 0.28 & 13\\
\hline
$4^+_1$ & 0.00 & 0.00 & 0.00 & 0.07 & 0.93 & & & 0.23 & 28 \\
$5^+_1$ & 0.00 & 0.00 & 0.01 & 0.08 & 0.85 & 0.07 & & 0.23 & 24 \\  
$6^+_2$ & 0.00 & 0.01 & 0.01 & 0.14 & 0.80 & 0.04 & 0.00 & 0.23 & 26\\
\end{tabular} 
\end{ruledtabular}
\end{table}

\begin{table}[t]
\caption{Same as TABLE~\ref{tab:s44} but for $^{43}$S.}
\label{tab:s43}
\begin{ruledtabular}
\begin{tabular}{rrrrrrrrrr}
$I^{\pi}_{\sigma}$  & \multicolumn{6}{c}{$|K|$} & $\beta$ & $\gamma$ \\ \cline{2-7}
 & 1/2 & 3/2 & 5/2 & 7/2 & 9/2 & 11/2 &  &  \\
\hline
$1/2^-_1$ & 1.00 & & & & & & 0.27 & 15\\
$3/2^-_1$ & 0.98 & 0.02 & & & & & 0.25 & 17 \\
$5/2^-_1$ & 0.97 & 0.03 & 0.00 & & & & 0.27 & 16 \\
$7/2^-_2$ & 0.96 & 0.04 & 0.00 & 0.00 & & & 0.25 & 16\\
$9/2^-_2$ & 0.96 & 0.04 & 0.00 & 0.00 & 0.00 & & 0.28 & 15\\
$11/2^-_1$ & 0.91 & 0.09 & 0.00 & 0.00 & 0.00 & 0.00 & 0.25 & 16\\
\hline
$7/2^-_1$ & 0.00 & 0.01 & 0.01 & 0.98 & & & 0.22 & 31 \\
$9/2^-_1$ & 0.00 & 0.01 & 0.04 & 0.95 & 0.01 & & 0.23 & 31 \\  
\hline
$11/2^-_2$ & 0.00 & 0.01 & 0.01 & 0.08 & 0.03 & 0.87 & 0.25 & 37\\
\end{tabular} 
\end{ruledtabular}
\end{table}

Supported by those very large overlaps, 
it is reasonable to deduce intrinsic properties 
of many-body wave functions from 
the corresponding  AM-VAP intrinsic states,   
$| \Phi(IM\sigma)\rangle$ in Eq.~(\ref{eq:amvap}). 
The intrinsic mass quadrupole moments $Q_0$ and $Q_2$ 
are given as the expectation values of the mass quadrupole operators 
$r^2 Y_{20}$ and $r^2 Y_{22}$ 
in $| \Phi(IM\sigma)\rangle$,
respectively, where the axes of the intrinsic frame are determined 
to diagonalize the quadrupole tensor and to satisfy the order   
$\langle Q_{zz}\rangle \geq \langle Q_{xx}\rangle \geq \langle
Q_{yy}\rangle$. It is noted that a similar method has been 
used to deduce shape fluctuation in exotic Ni isotopes 
from the Monte Carlo shell-model wave functions \cite{ni}. 
The quadrupole deformation parameters $(\beta, \gamma)$ are 
then defined in the usual way as $\beta = f_{\rm scale} 
\sqrt{\frac{5}{16\pi}} 
\frac{4\pi}{3R^2A}\sqrt{(Q_0)^2 + 2(Q_2)^2}$ 
and $\gamma=\arctan(\frac{\sqrt{2}Q_2}{Q_0})$, 
with $R=1.2A^{1/3}$~fm \cite{gogny_gcm}. 
Here, a factor $f_{\rm scale} = e_p/e+e_n/e$ is introduced 
to rescale quadrupole matrix elements between the shell-model space and 
the full single-particle space. 
Since the intrinsic axes are thus determined, the $K$ quantum number 
is well defined apart from its sign. 
The distribution of $K$, which is normalized to unity, is calculated 
by following the method shown in Ref. ~\cite{gogny_gcm}. 

The intrinsic properties of the AM-VAP states defined above are listed in
Table~\ref{tab:s44} and Table~\ref{tab:s43} 
for $^{44}$S and $^{43}$S, respectively. 
We first outline the properties of $^{44}$S. 
The $0^+_1$, $2^+_1$, $4^+_2$, $6^+_1$ sequence, connected with 
strong $E2$ matrix elements, is dominated by the $K=0$ state
as usually conceived for the ground-state band, whereas 
the $K=1$ component grows up with increasing spin 
because of the Coriolis coupling. 
The shape of the ground-state band evolves from triaxial to prolate. 
This shape evolution is very similar to the results of 
a beyond-mean-field calculation based on
the GCM \cite{s44gogny} and an analysis based on 
the quadrupole rotational invariants in the shell-model wave functions 
\cite{s44qinv}. 
On the other hand, the $4^+_1$, $5^+_1$, $6^+_2$ sequence, also 
connected with strong $E2$ matrix elements, is missing 
in the beyond-mean-field study of Ref.~\cite{s44gogny}, but appears 
in the shell model 
both with the SDPF-U interaction 
\cite{s44isomer, s44qinv} 
and with the SDPF-MU interaction \cite{sdpf-mu}.

The present AM-VAP calculation demonstrates that this band is 
strongly dominated by the $K=4$ state for the first time. 
Experimentally, while this $K$ assignment has been suggested in
\cite{s44isomer}, its basis is only the weak $E2$ transition to 
the $2^+_1$ level because the usual shell-model calculation cannot 
provide $K$ quantum numbers. In light nuclei, 
however, such a high-$K$ isomer has not been known 
and is rather unexpected. 
It is commonly believed that the high-$K$ isomerism occurs 
only in axially symmetric, stably deformed nuclei \cite{isomer_review}, 
whereas this condition is not satisfied in light nuclei. 
What is surprising in the present case is that 
the the concentration of the $K$ numbers takes place 
in the $K=4$ band in spite of the significant triaxiality 
of the $K=4$ states. 
The AM-VAP calculation shows that  
although the $K$ numbers are strongly mixed in the intrinsic 
state $|\Phi \rangle$ of Eq.~(\ref{eq:amvap}), 
the purity of $K$ is approximately restored 
after diagonalizing the Hamiltonian in the $K$ space. 

Why does the $K=4$ band emerge at such a low excitation energy 
in $^{44}$S? 
The structure of $^{43}$S brings key information on this question, 
which we briefly examine here.  
The energy levels in $^{43}$S are presented in
Fig.~\ref{fig:s-level} and their intrinsic properties are listed 
in Table~\ref{tab:s43}. 
The observed energy levels and the $E2$ transitions are well reproduced 
with the full shell-model and AM-VAP calculations. 
The strong $E2$ excitation to 
the 940 keV state \cite{s43e2}
and the isomeric state at 320 keV \cite{s43isomer, s43g},  
shown in Fig.~\ref{fig:s-level},  
suggest a possible coexistence of configurations. 
While the ground state should be deformed on the basis of the large  
$B(E2;{\rm g.s.} \to 940\,{\rm keV})$ value, the structure 
of $7/2^-_1$ has been less understood. 
Although an early analysis \cite{s43g} proposed a quasi-spherical state, 
the large quadrupole moment in this state measured later \cite{s43q}, 
$|Q| = 23(3)$~$e$fm$^2$, casts doubt on this interpretation. 
The AM-VAP analysis demonstrates 
that the ground state and the $7/2^-_1$ state are dominated 
by $K=1/2$ and $K=7/2$, respectively, and that this $K$ forbiddeness 
causes the isomerism in  $7/2^-_1$. 
Although this result is consistent with the suggestion of \cite{amd_s}, 
the present calculation is the first to quantitatively provide 
the distribution of $K$. 
The $K=7/2$ and $K=1/2$ states 
have triaxial and nearly prolate deformations, respectively, 
in accordance with \cite{s44qinv, amd_s}. Thus, 
similar to the $K=4$ band in $^{44}$S, the $K=7/2$ band in $^{43}$S 
has an approximately good $K$ number in spite of the development of 
triaxiality. 

Since $^{43}$S is an even-odd nucleus, the $K$ quantum number 
is determined by that of the unpaired neutron, $\Omega^{\pi}$. 
On the basis of the above analysis, we 
hereafter assume a good $\Omega^{\pi}$ 
quantum number even in the case of non axially-symmetric deformation.
As expected from the Nilsson diagram (see also Ref.~\cite{amd_s}),  
the orbit of the last neutron can be 
$\Omega^{\pi}=1/2^-$ which favors prolate deformation 
or $\Omega^{\pi}=7/2^-$ which favors oblate deformation.
Thus, the $K=1/2$ state favors a prolate shape because of the preference 
of prolate deformation for both the $\Omega^{\pi}=1/2^-$ neutron 
and the $^{42}$S core (see the potential energy surface of $^{42}$S, e.g., in
Ref.~\cite{sdpf-mu}). 
The $K=7/2$ state, on the other hand, is inclined to be triaxial 
due to the opposite shape preference between the last neutron and the
core. 

\begin{figure}[t]
 \begin{center}
\includegraphics[width=8.0cm,clip]{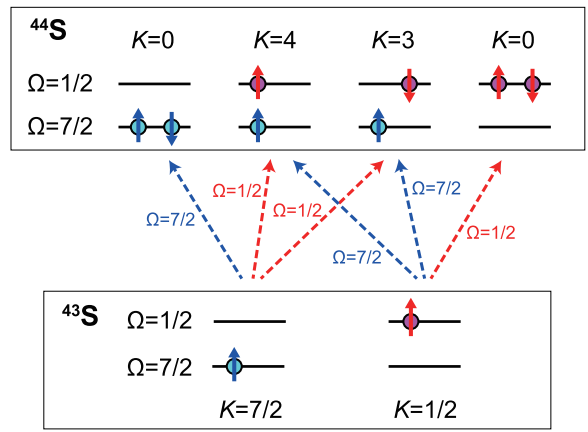}
 \caption{ 
(Color online)
Possible low-lying neutron configurations for $^{44}$S and $^{43}$S.  
}
 \label{fig:s-config}
 \end{center}
\end{figure}

Now we turn to $^{44}$S. It is reasonable 
to fix the $^{42}$S core and to take into account 
only the two quasiparticle degrees of freedom, 
$\Omega^{\pi}=1/2^-$ and $7/2^-$.
It is noted that the quasiparticle state is introduced, as usual,  
to include pairing correlation within the single-particle picture. 
Those two orbits 
are regarded to be nearly degenerate in this simple
picture similarly to 
the small energy difference between the $3/2^-_1$ and $7/2^-_1$ states 
in $^{43}$S.  
For the two-quasiparticle system of $^{44}$S, two $K=0$ states, 
one $K=3$ state, and one $K=4$ state can be constructed 
by adding a neutron as illustrated 
in Fig.~\ref{fig:s-config}. 
We then consider the competition of energy 
between different $K$ states 
by decomposing the energy of a state 
into the intrinsic energy and the rotational energy. 
Concerning the intrinsic energy, 
the lowest $K=0$ zero-quasiparticle state is in general lower than the 
$K \ne 0$ two-quasiparticle states 
even if the relevant quasiparticle states are degenerate 
because the former gains additional energy due to pairing,  
which amounts to $2\Delta$, where $\Delta$ is the pairing gap. 
Thus, the $K=0$ state usually dominates the yrast $0^+, 2^+, 4^+,
\ldots$ sequence. 
High-$K$ states are, however, advantageous over the $K=0$ state 
in terms of the loss of the rotational energy 
$E_{\rm rot} = (\hbar^2/2{\mathcal I})(I(I+1)-K^2) $, which is smaller 
for high-$K$
states. 
This is the standard picture for the occurrence of high-$K$ isomers  
seen usually in the medium-heavy-mass or heavy-mass regions
\cite{k_review}. 
In those heavier-mass regions, 
the $K \ne 0$ states cannot be lower than the $K=0$ member unless 
$K$ is sufficiently large because 
moments of inertia are rather large. 
On the other hand, for lighter nuclei with relatively small moments of 
inertia, a state with a modest $K$ can, in principle, intrude into the yrast 
$0^+, 2^+, 4^+,\ldots$ sequence 
if a high-$\Omega$ orbit is located very close to the Fermi surface.  
$^{44}$S is such a very rare case, 
but similar situations can occur in other nuclei.
For nuclei around $^{44}$S, two quasiparticle states are estimated to 
appear above $2\Delta
\approx 2.5$~MeV, where the pairing gap is evaluated from one-neutron 
separation energies of $^{43,44,45}$S. 
This energy estimate accounts for the excitation energy of the measured isomeric 
$4^+$ state. 
On the other hand, the $4^+$ state with $K=0$ is roughly estimated to 
lie around 3 MeV by assuming a normal $E_x(4^+)/E_x(2^+)$ ratio, 
$\sim 2.5$.  This is how the $K=4$ state becomes the yrast state in
$^{44}$S when $\Omega^{\pi}=7/2^-$ and $\Omega^{\pi}=1/2^-$ are 
nearly degenerate. 

In the beyond-mean-field approach of Ref.~\cite{s44gogny}, 
the low-lying $K=4$ state is missing 
despite many other similarities, 
such as shapes, to the present calculations. 
This is due to the restriction of time-reversal symmetry 
on the intrinsic states imposed in the calculation of Ref.~\cite{s44gogny}. 
The AM-VAP calculation in fact  
demonstrates that the time-reversal symmetry is almost completely broken 
for the intrinsic state of the $K=4$ state. 
The overlap probabilities between $| \Phi \rangle$ 
of Eq.~(\ref{eq:amvap}) and its time-reversed 
state ${\mathcal T} | \Phi \rangle$ are calculated to be only 0.02-0.08
for the $K=4$ members.
The breaking of time-reversal symmetry in the intrinsic state 
is almost solely attributed to the neutron part of the wave function.  

Here we confirm that the actual shell-model wave function of $4^+_1$ 
is indeed dominated by the above-mentioned two-quasiparticle state by 
calculating its spectroscopic strengths.
The full shell-model calculation leads to 
large overlap probabilities of the $4^+_1$ state 
with the ${\mathcal A}[\nu p_{3/2} \otimes 7/2^-_1]^{J=4}$ 
and ${\mathcal A}[\nu f_{7/2} \otimes 7/2^-_2]^{J=4}$ states 
(0.66 and 0.54, respectively), where 
$\mathcal A$ denotes antisymmetrization and normalization.
On the other hand, its overlap probabilities with 
${\mathcal A}[\nu f_{7/2} \otimes 7/2^-_1]^{J=4}$ and 
${\mathcal A}[\nu p_{3/2} \otimes 7/2^-_2]^{J=4}$ are much smaller 
(0.19 and 0.00, respectively). 
This is a direct consequence of the dominance of 
the configurations illustrated in Fig.~\ref{fig:s-config}, 
since  the $7/2^-_1$ and $7/2^-_2$ states are dominated by 
$K=7/2$ and $K=1/2$, respectively, and the $\Omega^{\pi}=1/2^-$ and 
$\Omega^{\pi}=7/2^-$ single-particle states are 
dominated by $p_{3/2}$ and $f_{7/2}$, respectively.

We also point out that the configurations shown in 
Fig.~\ref{fig:s-config} account for the reason why the $0^+_2$ 
state in $^{44}$S lies extraordinary low. 
In Fig.~\ref{fig:s-config}, two $K=0$ states exist, having 
similar diagonal energies. 
The off-diagonal Hamiltonian matrix element between those 
$K=0$ states strongly mixes and repels each other, but 
the excited $K=0$ state can be rather low if the two quasiparticle
states are nearly degenerate.
The strong mixing between those two $K=0$ states is supported 
by the shell-model result that 
the overlap probability of the $0^+_1$ state 
with ${\mathcal A}[\nu f_{7/2} \otimes 7/2^-_1]^{J=0}$ (0.61) 
is close to 
the one with ${\mathcal A}[\nu p_{3/2} \otimes 3/2^-_1]^{J=0}$ (0.39). 
When the total spin for each $K=0$ state increases, 
the state with a larger moment of inertia 
is relatively lowered. Here, 
the $K=0$ state that occupies $\Omega^{\pi}=1/2^-$ is the case 
because of larger, prolate deformation.  
This accounts for the shape evolution toward 
prolate deformation within the $K=0$ band shown in Table~\ref{tab:s44}. 
It is noted that the ground state and $K=4$ states deviate from 
a prolate shape 
because they occupy the oblate-favored $\Omega^{\pi}=7/2^-$ orbit. 

Finally, we note that the $K=3$ state illustrated 
in Fig.~\ref{fig:s-config} is also found in the shell-model 
and AM-VAP calculations. In the shell-model
calculation, 
the $3^+_1$ level appears very close to $4^+_1$, 
dominated by the $K=3$ AM-VAP state. 
The $K=3$ states are distinguishable from the $K=4$ states 
by neutron intrinsic-spin expectation value 
because of difference in neutron spin orientation. 

In conclusion, we have clarified that the isomeric $4^+_1$ state 
in $^{44}$S observed recently \cite{s44isomer} originates from the
dominance of the $K=4$ state by means of the 
variation after angular-momentum projection (AM-VAP) approximation to the 
shell model. 
This result is very robust because it 
is understood within a unified picture 
about the occurrence of exotic isomeric states in $^{43,44}$S 
including the $7/2^-_1$ state in $^{43}$S and the $0^+_2$ in $^{44}$S. 
The $K=4$ state is missing in the beyond-mean-field
calculation \cite{s44gogny} due to the restriction of time-reversal 
symmetry and may appear without it. 
Having extraordinary light mass and triaxial deformation, 
the $K=4$ isomer in $^{44}$S is distinct from previously known 
high-$K$ isomers. Thus, 
the possibility of the occurrence of high-$K$ isomerism 
is greatly extended to the whole 
chart of nuclides, which 
provides new experimental and theoretical opportunities. 

Y.~U. thanks Prof. I. Wiedenh\"over for his enlightening discussions 
at an early stage 
and Dr. T.~Shizuma for useful comments on high-$K$ isomers. 
The conventional shell-model calculations were performed 
with the code {\sc mshell64} \cite{mshell64}, and 
the AM-VAP calculations were carried out with the advanced 
Monte-Carlo shell-model code \cite{rmcsm}.
This work was supported in part by JSPS KAKENHI Grant Numbers 20244022, 
21740204, and 25870168. 
This work has been supported by HPCI (hp120284 and hp130024), and 
is a part of the RIKEN-CNS joint research project on large-scale
nuclear-structure calculations.



\begin{thebibliography}{}

\bibitem{bohr}
For instance, A.~Bohr and B.~R.~Mottelson, 
{\it Nuclear Structure}, Vol. II, Benjamin, New York, 1975.

\bibitem{s44isomer}
D.~Santiago-Gonzalez {\it et al.}, Phys. Rev. C {\bf 83}, 
061305(R) (2011). 

\bibitem{s44gogny}
T.~R.~Rodr\'iguez and J.~L.~Egido, Phys. Rev. C {\bf 84}, 
051307(R) (2011). 

\bibitem{s44qinv}
R.~Chevrier and L.~Gaudefroy, Phys. Rev. C {\bf 89}, 
051301(R) (2014). 

\bibitem{s44be2}
T.~Glasmacher {\it et al.}, Phys. Lett. B {\bf 395}, 163 (1997). 

\bibitem{s44zero1}
S.~Gr\'evy {\it et al.}, Eur. Phys. J. A {\bf 25}, 111 (2005). 

\bibitem{s44zero2}
C.~Force {\it et al.}, Phys. Rev. Lett. {\bf 105}, 102501 (2010). 

\bibitem{s43isomer}
F.~Sarazin {\it et al.}, Phys. Rev. Lett. {\bf 84}, 5062 (2000). 

\bibitem{s43g}
L.~Gaudefroy {\it et al.}, Phys. Rev. Lett. {\bf 102}, 092501 (2009); 
P.F.~Mantica, Physics {\bf 2}, 18 (2009). 

\bibitem{sdpf-u}
F.~Nowacki and A.~Poves, Phys. Rev. C {\bf 79}, 014310 (2009). 

\bibitem{n28sm}
L.~Gaudefroy, Phys. Rev. C {\bf 81}, 064329 (2010). 

\bibitem{epqqm}
K.~Kaneko, Y.~Sun, T.~Mizusaki, and M.~Hasegawa, 
Phys. Rev. C {\bf 83}, 014320 (2011). 

\bibitem{sdpf-mu}
Y.~Utsuno, T.~Otsuka, B.~A.~Brown, M.~Honma, T.~Mizusaki, and 
N.~Shimizu, Phys. Rev. C {\bf 86}, 051301(R) (2012).

\bibitem{peru}
S.~P\'eru, M.~Girod, and J.~F.~Berger, 
Eur. Phys. J. A {\bf 9}, 35 (2000). 

\bibitem{n28gogny}
R.~Rodr\'iguez-Guzm\'an, J.~L.~Egido, and L.~M.~Robledo, 
Phys. Rev. C {\bf 65}, 024304 (2002). 

\bibitem{ddpc1}
Z.~P.~Li, J.~M.~Yao, D.~Vretenar, T.~Nik\v{s}i\'c, H.~Chen, 
and J.~Meng, Phys. Rev. C {\bf 84}, 054304 (2011). 

\bibitem{amd_s}
M.~Kimura, Y.~Taniguchi, Y.~Kanada-En'yo, H.~Horiuchi, and K.~Ikeda, 
Phys. Rev. C {\bf 87}, 011301(R) (2013). 

\bibitem{isomer_chart}
P.~M.~Walker, AIP Conf. Proc. {\bf 819}, 16 (2006). 

\bibitem{isomer_review}
P.~M.~Walker and G.~D.~Dracoulis, Nature {\bf 399}, 35 (1999). 

\bibitem{ring}
P.~Ring and P.~Schuck, {\it The Nuclear Many-Body Problem}, 
Springer-Verlag, Berlin, 1980. 

\bibitem{numerical}
W.~H.~Press, S.~A.~Teukolsky, W.~T.~Vetterling, 
and B.~P.~Flannery, {\it Numerical Recipes in FORTRAN 77: 
The Art of Scientific Computing}, Second Edition, 
Cambridge University Press, Cambridge, 1992. 

\bibitem{rmcsm}
N.~Shimizu, T.~Abe, Y.~Tsunoda, Y.~Utsuno, T.~Yoshida, T.~Mizusaki, 
M.~Honma, and T.~Otsuka, Prog. Theor. Exp. Phys. {\bf 2012}, 
01A205 (2012). 

\bibitem{s44gamma}
L.~C\'aceres {\it et al.}, Phys. Rev. C {\bf 85}, 024311 (2012). 

\bibitem{s43e2}
R.~W.~Ibbotson, T.~Glasmacher, P.~F.~Mantica, and H.~Scheit, 
Phys. Rev. C {\bf 59}, 642 (1999). 

\bibitem{ni}
Y.~Tsunoda, T.~Otsuka, N.~Shimizu, M.~Honma, and Y.~Utsuno, 
Phys. Rev. C {\bf 89}, 031301(R) (2014). 

\bibitem{gogny_gcm}
T.~R.~Rodr\'iguez and J.~L.~Egido, Phys. Rev. C {\bf 81}, 
064323 (2010). 

\bibitem{s43q}
R.~Chevrier {\it et al.}, Phys. Rev. Lett. {\bf 108}, 162501 (2012). 

\bibitem{k_review}
For instance, P.~M.~Walker and G.~D.~Dracoulis, 
Hyperfine Int. {\bf 135}, 83 (2001). 

\bibitem{mshell64}
T.~Mizusaki, N.~Shimizu, Y.~Utsuno, and M.~Honma, {\sc mshell64} code 
(unpublished). 

\end{thebibliography}
\end{document}